\documentclass[twocolumn,aps,pra,superscriptaddress,floatfix]{revtex4}

\usepackage[english]{babel}
\usepackage{blindtext}
\usepackage{siunitx}
\usepackage{float}
\usepackage{xcolor}
\usepackage{adjustbox}
\usepackage{graphicx}
\usepackage{dcolumn}
\usepackage{bm}
\usepackage{kantlipsum}
\PassOptionsToPackage{hyphens}{url}\usepackage{hyperref}
\usepackage[mathlines]{lineno}
\usepackage{physics}
\usepackage{cleveref}

\begin{document}
\title{Quantum-enhanced biosensing enables earlier detection of bacterial growth}

\author{Rayssa B. de Andrade}
\affiliation{Center for Macroscopic Quantum States bigQ, Department of Physics, Technical University of Denmark, Fysikvej 307, DK-2800 Kgs.\ Lyngby, Denmark.}
\author{Anne Egholm H\o gh}
\affiliation{Department of Health Technology Magnetic Resonance, Technical University of Denmark, Ørsteds Plads, 349, DK-2800 Kgs.\ Lyngby, Denmark.}
\author{Gaetana Spedalieri}
\affiliation{Department of Computer Science, University of York, York YO10 5GH, United Kingdom.}
\author{Stefano Pirandola}
\affiliation{Department of Computer Science, University of York, York YO10 5GH, United Kingdom.}
\author{Kirstine Berg-S\o rensen}
\affiliation{Department of Health Technology Magnetic Resonance, Technical University of Denmark, Ørsteds Plads, 349, DK-2800 Kgs.\ Lyngby, Denmark.}
\author{T.\,Gehring}
\affiliation{Center for Macroscopic Quantum States bigQ, Department of Physics, Technical University of Denmark, Fysikvej 307, DK-2800 Kgs.\ Lyngby, Denmark.}
\author{Ulrik L. Andersen}
\affiliation{Center for Macroscopic Quantum States bigQ, Department of Physics, Technical University of Denmark, Fysikvej 307, DK-2800 Kgs.\ Lyngby, Denmark.}
\email{ulrik.andersen@fysik.dtu.dk}

\begin{abstract}
Rapid detection of bacterial growth is crucial in clinical, food safety, and environmental contexts, yet conventional optical methods are limited by noise and require hours of incubation. Here, we present the first experimental demonstration of a quantum-enhanced photometric measurement for early bacterial detection using squeezed light. By monitoring the optical absorbance of an Escherichia coli culture with a quantum probe, we achieve a sensitivity beyond the shot-noise limit, enabling identification of growth onset up to 30 minutes earlier than with a classical sensor. The noise reduction is validated through statistical modeling with a truncated Gaussian distribution and hypothesis testing, confirming earlier detection with low false-alarm rates. This work illustrates how quantum resources can improve real-time, non-invasive diagnostics. Our results pave the way for quantum-enhanced biosensors that accelerate detection of microbial growth and other biological processes without increasing photodamage.
\end{abstract}

\maketitle
\section{Introduction}

The rapid and accurate detection of bacterial growth is critical across a wide range of domains, including clinical diagnostics, food safety, pharmaceutical quality control, and environmental monitoring~\cite{zhao2024,marro2022}. In many of these settings, the earliest possible identification of microbial contamination can significantly improve treatment outcomes, prevent the spread of infection, or reduce economic loss. Yet, conventional optical methods for detecting bacterial growth, such as turbidity-based absorbance measurements, are fundamentally limited in their sensitivity and speed. These methods typically rely on macroscopic accumulation of biomass, which can delay detection by hours after bacterial proliferation has begun~\cite{Mcgoverin2021}.

A key challenge in early-stage bacterial detection is the inherently low signal-to-noise ratio encountered when monitoring sparse cell populations \cite{Hu2017,Locke2020}. One straightforward way to improve sensitivity is to increase optical probe power, but this is constrained by the risk of photodamage or heating, especially for living cells and fragile biological samples~\cite{Dubourg2018}. This trade-off between sensitivity and invasiveness is a pervasive limitation of classical optical sensing, especially in the low-photon-flux regime necessary for non-destructive biological measurements~\cite{fu_characterization_2006, galli_intrinsic_2014, talone_phototoxicity_2021,Belkum2020}.

Quantum-enhanced optical measurements offer a promising alternative. By exploiting non-classical states of light, such as squeezed states, it is possible to suppress quantum noise below the shot-noise limit, enabling more sensitive detection without increasing probe power. This quantum advantage has, for example, been demonstrated in absorption~\cite{brida2010} and Raman~\cite{andrade_quantum-enhanced_2020,xu_stimulated_2022} spectroscopy, and in particular in a biological context~\cite{taylor_biological_2013,casacio_quantum-enhanced_2021}. Crucially, these quantum resources allow for information-rich measurements while minimizing the energetic footprint on the sample which is an essential feature for living systems.

In this work, we demonstrate a quantum-enhanced approach to bacterial growth detection using squeezed light and a hypothesis testing analysis algorithm. By monitoring the optical absorbance of a bacterial culture over time with both classical (coherent) and quantum (squeezed) probes, we show that the onset of growth can be identified significantly earlier when using the quantum-enhanced measurement. In fact, the squeezed-light sensor detected Escherichia coli growth up to approximately 30 minutes earlier than an equivalent classical optical sensor. This advance illustrates the potential of quantum sensing for real-time, low-invasiveness biosensing and paves the way for future quantum-enabled diagnostics tailored to sensitive biological materials.

\section{Experimental setup and results}

\begin{figure*}[htpb]
\centering
\includegraphics[width =\textwidth]{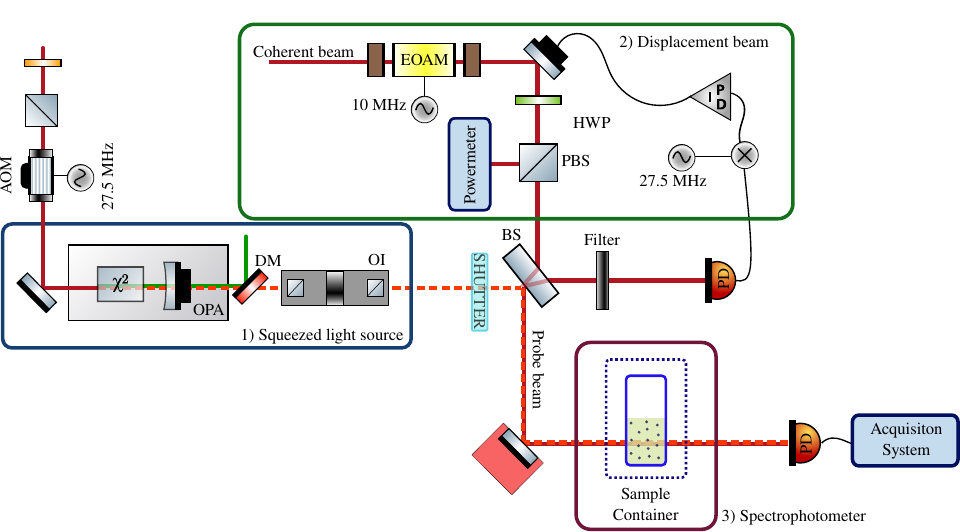}
\caption{Experimental setup. The system contains three main parts: 1) Squeezed light source, 2) Displacement beam module, and 3) Photometer. OPA: Optical Parametric Amplifier, DM: Dichroich mirror, OI: Optical isolator, HWP: Half-wave plate, PBS: Polarizing beam-splitter, BS: Beam-splitter, PD: Photodetector, EOAM: Electro-optic modulator, AOM: Acoustic optical modulator.}
\label{fig:Experimental_setup}
\end{figure*}

{\bf Quantum-Enhanced photometer setup}-- 
To realize a quantum-enhanced biosensor, we developed a photometric system that combines a bright squeezed-light probe with a temperature-controlled micro-incubator for live bacterial monitoring (Fig.~\ref{fig:Experimental_setup}). The setup comprises three functional modules: a squeezed-light source, a displacement stage that converts the squeezed vacuum into a bright probe, and a photometer that transmits the probe through a bacterial sample and records its transmission in real time. 

The source of squeezed states of light is a doubly resonant optical parametric amplifier (OPA) consisting of a periodically poled KTP (PPKTP) crystal in an optical cavity (see supplementary material for more details). When pumped below threshold, the OPA produces a squeezed vacuum state of light. We measure a squeezing level (vacuum noise reduction) of \SI{-6.19}{\decibel} at \SI{1064}{\nano\meter} on a homodyne detector at \SI{10}{\mega\hertz} sideband frequency.

The squeezed vacuum is converted into a bright squeezed probe beam via a coherent displacement. In the displacement module, the squeezed vacuum state interferes with a coherent state at a highly asymmetric beam splitter with a splitting ratio of 99:1 (with 99\,\% transmission for the squeezed beam to minimize loss of squeezing). The coherent state (derived from the same Nd:YAG laser) is amplitude modulated at \SI{10}{\mega\hertz} using an electro-optic amplitude modulator (EOAM) to produce a faint intensity modulation that serves as the signal. The coherent carrier served as the local oscillator and phase reference for the squeezed beam. Phase stabilization was achieved by demodulating the interference between the two beams and feeding back to a piezoelectric actuator (details in supplementary material). The error signal was processed by a proportional-integral controller whose output actuated a piezo-electric phase shifter in the coherent beam path, maintaining a stable phase lock. The power of the coherent beam was controlled with a half-waveplate and a polarizing beam splitter; it was set to \SI{1.7}{\milli\watt} (measured after the 99:1 beam splitter) for all measurements.

The photometer is a home-built system that consists of a mini-incubator (air-cooled, Peltier-controlled single cuvette holder from Mountain Photonics (QNW qX2)) that holds the cuvette containing the sample under investigation, keeping it at \SI{37}{\celsius} while stirring at 250 rpm. The mini-incubator has a clear aperture (\SI{12}{\milli\meter} high x \SI{10}{\milli\meter} wide) through which the probe light can pass the sample. The signal output on the other side of the incubator is detected by a resonant photodetector at \SI{10}{\mega\hertz}. This home-built detector has a demodulation system coupled to it, and the down-mixed signal is low-pass filtered at \SI{25}{\kilo\hertz} and acquired using a data acquisition card (NI PCIe-6323, X Series DAQ). A custom software was developed to perform the data acquisition and analysis.

{\bf Measurement procedure} -- Before introducing bacteria, we characterized losses and baseline noise in the system. With no cuvette in place (open beam path), the detected squeezing level was \SI{-3.98}{\decibel}, lower than the source squeezing (–6.2\,dB) due to optical loss in the path. We then inserted a sterile cuvette containing \SI{3}{\milli\liter} of growth medium (Luria-Bertani broth with no bacteria) to measure the “blank” transmission $\eta_{\mathrm{blank}}$ of the medium. This blank measurement accounts for loss due to the cuvette and medium (we found $\eta_{\mathrm{blank}}\approx0.80$, i.e. about 20\% loss from the medium and cuvette).

For the bacterial sample, we used Escherichia coli (strain MG1655) as a model organism. A starter culture was grown in LB broth at 37°C overnight (24h) to reach a high optical density. At time $t=0$ of the experiment, we inoculated the photometer’s cuvette (containing fresh sterile LB medium) with a small aliquot of the overnight culture (diluting it to a low concentration). We continuously collected data for 5 hours, taking sequential measurements approximately every minute. This 5-hour measurement duration covers the full bacterial growth cycle,
from the initial lag phase to exponential proliferation. At each time point, two consecutive 5s data acquisitions were performed: first using a classical coherent state probe (with the squeezed beam shuttered off), and immediately after using the squeezed-state probe (with the squeezed beam unshuttered and phase-lock engaged). Both measurements used identical optical power (1.7 mW) and were sampled at 250 kHz. In parallel, we recorded the transmitted DC power of the probe (the coherent carrier) to correct for any slow drifts in laser intensity.

\begin{figure}[hbt!]
\centering
\includegraphics[width=1\linewidth]{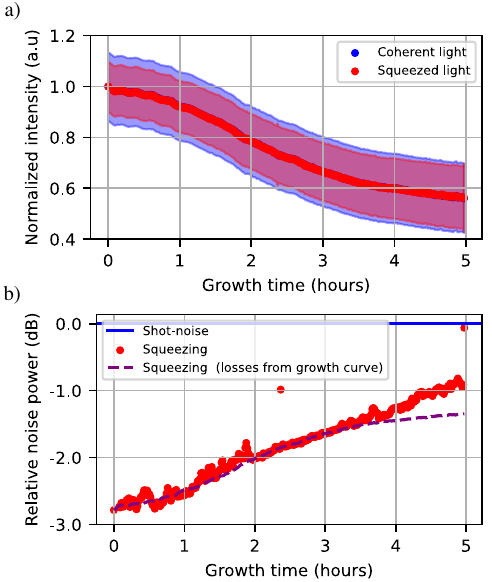}
\caption{Measurement dataset: red dots indicate the squeezed state and blue dots the coherent state of light. (a) Normalized mean value (in arbitrary units) as a function of time (hours). (b) Variance of the squeezed and coherent states, normalized to the shot-noise level and expressed in dB. The purple dashed curve shows the expected squeezing level based on losses inferred from the growth curve.}
\label{fig:stats_Ecoli}
\end{figure}

{\bf Growth curve results} -- 
Figure~\ref{fig:stats_Ecoli} shows a representative dataset from a 5-hour growth measurement. Panel (a) plots the normalized transmitted intensity (relative to the initial intensity) as a function of time for both the coherent-light measurement (blue circles) and the squeezed-light measurement (red squares). Each data point represents the mean intensity over a 5s acquisition, and the shaded regions indicate the ±1 standard deviation range of intensity fluctuations during that period. As expected, the mean transmitted intensity gradually decreases over time as the bacteria multiply in the medium, increasing the sample’s absorbance (optical loss). Correspondingly, the measured noise level with the squeezed probe increases over time relative to its initial value. In the early stages (low absorbance), the squeezed light maintains a reduced noise variance well below the shot-noise level, but as absorbance grows, the quantum advantage is eroded and the noise approaches the shot-noise limit. This behavior is quantified in Fig.~\ref{fig:stats_Ecoli}(b), which plots the variance of the intensity measurement (in dB relative to shot-noise) for the squeezed (red) and coherent (blue) cases over time. Initially, about 3\,dB of noise reduction is observed with the squeezed state, and this advantage diminishes as the transmissivity drops. The purple trace in Fig.~\ref{fig:stats_Ecoli}(b) shows the expected squeezing level as a function of time, calculated from the measured optical loss (absorbance) at each point. We see that the observed noise closely follows this expected trend. The occasional outlier points in Fig.~\ref{fig:stats_Ecoli}(b) correspond to moments when the phase lock of the squeezed light was temporarily lost (resulting in a brief loss of squeezing); these points were excluded from further analysis.

\begin{figure}[htb!]
\centering
\includegraphics[width=1\linewidth]{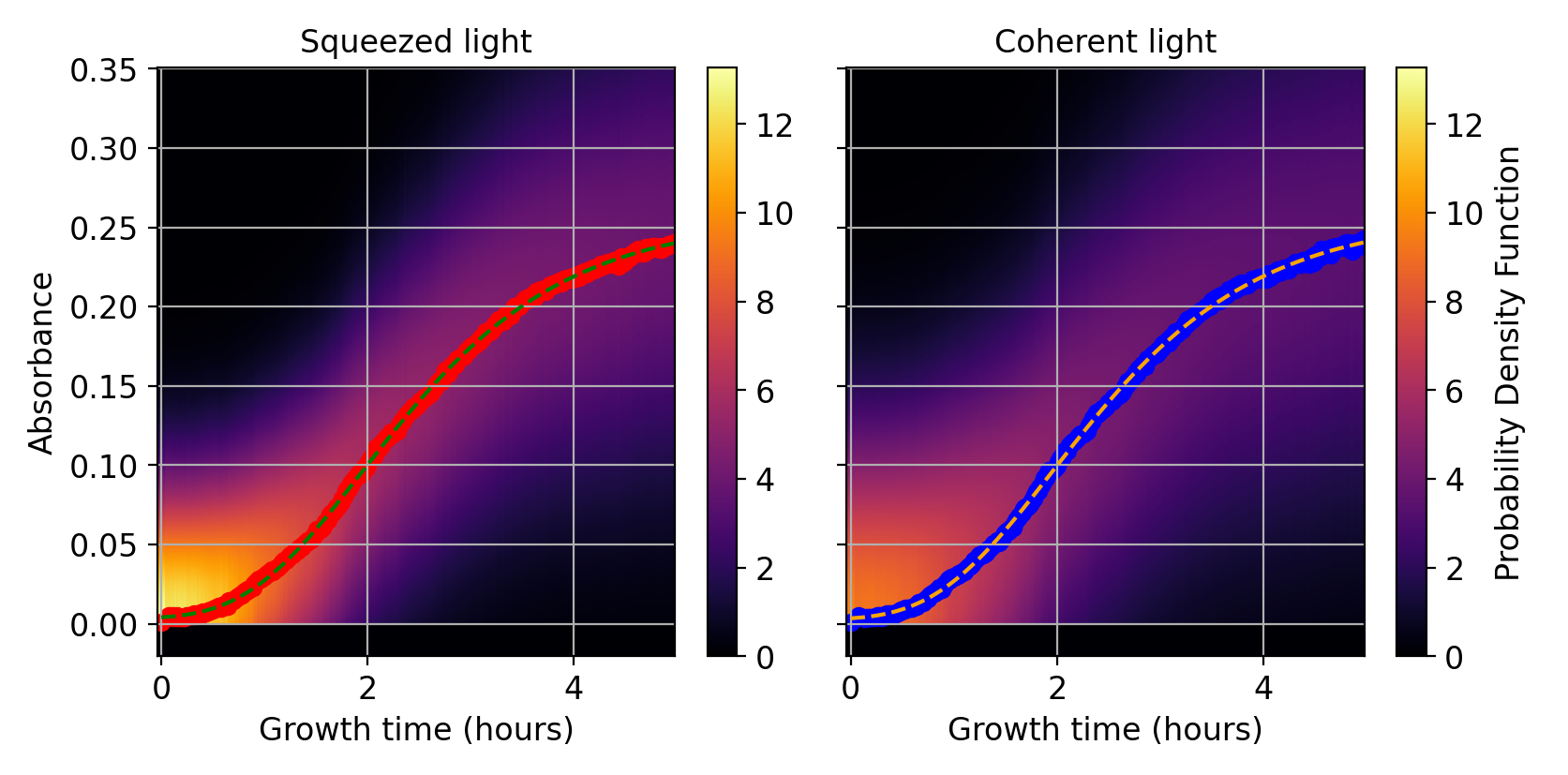}
\caption{Probability density function. Growth curve of E. coli in terms of absorbance versus growth time (hours). The data are fitted with a Gompertz function given by Eq. (2). The red and blue curves represent the mode of the truncated distribution for measurements performed with squeezed and coherent states of light, respectively.}
\label{fig:OD_GT_paper}
\end{figure}

From the raw transmission, we calculate the sample's absorbance, $A(t)$ as a function of time: $A = - \log_{10} \eta$, where $\eta$ is the transmissivity of the sample, calculated as $\eta = \eta_\text{sample}/\eta_\text{blank}$. 
At $t=0$, $A \approx 0$ (no excess absorption beyond the blank). As bacteria grow and increase light scattering/absorption, $A(t)$ rises. Figure \ref{fig:OD_GT_paper} presents the growth of absorbance over time, along with a statistical model fit. In converting the data to absorbance, one complication arises: at very early times, the measured absorbance values are extremely low and within the noise, leading to occasional negative absorbance estimates (which are non-physical). To account for this, we model the measurement noise in $A$ as a truncated Gaussian distribution \cite{spedalieri_detecting_2020} that is confined to $A \ge 0$. Following Ref.~\cite{spedalieri_detecting_2020}, the truncated normal distribution yields adjusted moments given by:
\begin{equation}
    \bar{A}^{'} = \bar{A} + g(\omega)\sigma_{A}, \; \; \sigma_{A}^{'} = \sigma_{A}\sqrt{1 + \omega g(\omega)-g(\omega)^{2}}
    \label{mean}
\end{equation}
where $\bar{A}$ and $\sigma_A$ are the mean and standard deviation of the uncorrected (symmetric) distribution, $\omega = -\bar{A}/\sigma_A$, and $g(\omega) = \frac{2 \mathcal{N}(\omega)}{1 - \operatorname{erf}(\omega/\sqrt{2})}$. Here $\mathcal{N}(x)$ is the standard normal distribution density at $\omega$ and $\operatorname{erf}(x)= 2\pi^{-1/2}\int_{0}^{x}e^{-x^{2}}dx$ is the error function. Equation (\ref{mean}) provides the mean $\bar{A}'$ and standard deviation $\sigma'_A$ of the absorbance, corrected for the fact that $A$ cannot drop below 0. Using this model, we obtain a physically meaningful probability distribution for $A$ at each time point in our experiment. 

In Fig. \ref{fig:OD_GT_paper}, we plot the truncated Gaussian probability density of the absorbance as a function of time (lighter regions correspond to higher probability density).  The solid lines in Fig. \ref{fig:OD_GT_paper}  trace the mode (peak value) of the absorbance distribution at each time for the squeezed-light data (red curve) and the coherent-light data (blue curve). These mode curves represent the most likely absorbance (or equivalently, the smoothed growth curves) extracted from the noisy data. To quantitatively characterize the growth, we fit each curve with a four-parameter Gompertz model \cite{tjorve_use_2017} which is a standard sigmoid model for bacterial growth \cite{mira_estimating_2022, krishnamurthi_new_2021}. The Gompertz function is given by
\begin{equation}
    A(t) = a\cdot e^{-\exp(\frac{\mu \cdot e}{a}(\theta - t) + 1)} + A_{bk},
    \label{eq:gompertz}
\end{equation}
where $a$ is the asymptotic absorbance, i.e. the absorbance when $t \rightarrow \infty$, $\mu$ is the growth rate in the mid-log phase, $\theta$ is the lag time before growth ramps up, and $A_{bk}$ is a constant offset accounting for any background absorbance of the blank sample. We fit Eq.~(\ref{eq:gompertz}) to the mode of each dataset (coherent and squeezed) using a nonlinear least-squares algorithm. Table \ref{tab:gompertz} summarizes the best-fit parameters associated with the two cases. Notably, the fitted growth parameters are virtually the same for the coherent and squeezed measurements: for example, both yield a growth rate $\mu \approx 0.0835~\text{h}^{-1}$ and lag $\theta \approx 0.84~\text{h}$. This implies that, as expected, the presence of squeezing does not perturb the biological process; it simply provides a more sensitive readout. The fitted $\mu$ corresponds to an approximate bacterial doubling time of $\ln(2)/\mu \approx 3.6$\,hours, consistent with typical Escherichia coli growth in rich medium at 37\,$^o$C using a small cuvette. We also observe that the variance in the absorbance is smaller for the squeezed-light data in the early growth phase: in Fig.~\ref{fig:OD_GT_paper}, the red (quantum) curve has tighter uncertainty bounds than the blue (classical) curve for $t \lesssim 2$\,h. This reduction of uncertainty in the initial regime is the quantum-enhanced advantage that we leverage for earlier detection, as described next. 
\begin{table}
\centering
\begin{adjustbox}{width=\columnwidth, center}
\begin{tabular}{ccccc}
\hline
& $a$ & $\mu [1/h]$ & $\theta [h]$ & $A_{bk}$ \\
\hline
Coherent states & $0.2546 \pm 0.0008 $ & $0.0835 \pm 0.0003$ & $ 0.837 \pm 0.007 $ & $0.0026 \pm 0.0004$ \\
Squeezed states & $0.253 \pm 0.001$ & $0.0835 \pm 0.0002$ & $ 0.839 \pm 0.007$ & $ 0.0033 \pm 0.0003$  \\
\hline
\end{tabular}
\end{adjustbox}
\caption{Fitted Gompertz growth parameters for E. coli absorbance curves. a: asymptotic absorbance; $\mu$: growth rate; $\theta$: lag time; A$_{bk}$: background absorbance.}
\label{tab:gompertz}
\end{table}
\section{Early detection}

While both the coherent and squeezed probes ultimately detect bacterial growth, the quantum-enhanced probe identifies the onset of growth earlier, particularly during the initial 2–3 hours. The advantage is most pronounced in the first 2–3 hours, after which the measurement noise with squeezed light approaches the shot-noise limit and the benefit diminishes (see Fig.~\ref{fig:stats_Ecoli}(b)). To rigorously quantify this temporal advantage, we apply a statistical hypothesis-testing framework that estimates how early a quantum probe can distinguish bacterial growth from the noise-limited baseline.
\begin{figure}[htb!]
\centering
\includegraphics[width=1\linewidth]{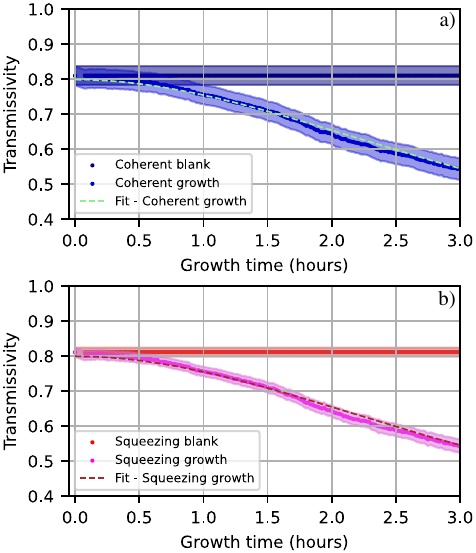}
\caption{Transmissivity as a function of growth time (hours). In both panels, the constant curve corresponds to the transmissivity of the blank measurement, while the polynomial curve represents the growth process. (a) Coherent-state measurements are shown in blue. (b) Squeezed-state measurements are shown in red. In both cases, the growth curve becomes distinguishable from the blank approximately half an hour earlier when using squeezed states of light.}
\label{fig:eta_paper}
\end{figure}
Our goal is to determine, for each probe, the earliest time at which we can reliably discriminate a growing sample from a blank (no-growth) sample. We adopt a binary hypothesis testing framework \cite{spedalieri_detecting_2020}. The null hypothesis $H_0$ corresponds to no growth, meaning the sample’s transmissivity remains at the blank level $\eta_{\mathrm{blank}}$ (apart from noise). The alternative hypothesis $H_1$ corresponds to growth at a given time $t > 0$, meaning the true transmissivity has fallen to $\eta(t) < \eta_{\mathrm{blank}}$ due to bacterial presence. We assume that under each hypothesis, the estimator $\hat{\eta}$ (our measured transmissivity) follows a Gaussian distribution: under $H_0$, $\hat{\eta}$ is centered at $\eta_{\mathrm{blank}}$ with standard deviation $\sigma_{\eta_{bk}}$ (determined by the measurement noise, which in our case includes shot noise and technical noise); under $H_1$, $\hat{\eta}$ is centered at $\eta(t)$ with standard deviation $\sigma_{\eta(t)}$. The measurement variance $\sigma^2_{\eta}$ differs between the two probes; the squeezed-light probe exhibits consistently lower noise (Fig.~\ref{fig:stats_Ecoli}). Given these distributions, one can set a decision threshold $\tau$ on the measured transmissivity to decide growth vs. no-growth. Two types of errors can occur: a false positive (false alarm), where we conclude growth while actually $H_0$ is true; and a false negative (missed detection), where we fail to detect growth even though $H_1$ is true. The probabilities of these errors depend on $t$ and $\tau$ and are given by integrals of the Gaussian tails. Following the derivations in Refs.~\cite{spedalieri_detecting_2020,spedalieri_optimal_2021}, we write the false-positive probability as a function of threshold $\tau$:
\begin{align}
    p_{FP}(\tau) = \frac{1}{2\mathcal{N}_{0}}\left\{\operatorname{erf}\left[ \frac{\eta_{bk}}{\sqrt{2}\sigma_{\eta_{bk}}} \right] - \operatorname{erf}\left[ \frac{\eta_{bk}- \tau}{\sqrt{2}\sigma_{\eta_{bk}}} \right]\right\} \label{pFP} \\
    p_{FN}(\tau, t) = \frac{1}{2\mathcal{N}_{1}}\left\{\operatorname{erf}\left[ \frac{\eta(t)-\tau}{\sqrt{2}\sigma_{\eta(t)}} \right] - \operatorname{erf}\left[ \frac{\eta(t)- 1}{\sqrt{2}\sigma_{\eta(t)}} \right]\right\}
    \label{pFN}
\end{align}
where the normalization factor is $\mathcal{N}_{i} = \int_{0}^{1} p_{i}(\eta)d\eta$.

\begin{figure}[htb!]
\centering
\includegraphics[width=1\linewidth]{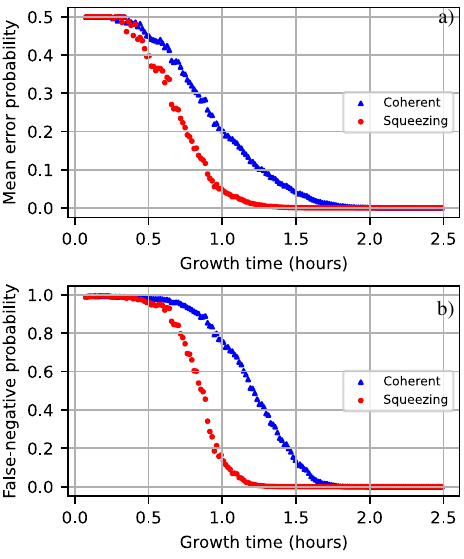}
\caption{(a) Mean error probability as a function of growth time (hours). (b) False-negative probability as a function of growth time (hours). Red curves represent measurements performed using squeezed states of light; blue curves correspond to measurements using coherent states.}
\label{fig:func_prob_paper}
\end{figure}

In practice, $\eta_{bk}$ is close to 1 after normalization (for our blank, $\eta_{bk} = 0.799$ before normalization, which we set as the reference 1.0). These expressions reduce to the familiar Gaussian error-function forms when transmissivity is unbounded; the additional terms account for the physical constraint that $\eta$ cannot exceed unity. Here the terms account for the fact that transmissivity cannot exceed 1. We evaluate two detection criteria for each probe type: asymmetric and symmetric tests. In an asymmetric test, one decides on a tolerable false-alarm probability (false-positive rate) and then determines how the false-negative probability evolves over time with that fixed threshold. For example, one might require $p_{\text{FP}}(\tau) = 0.001$ (0.1\% false alarms) and ask at what time $t$ the false-negative probability $p_{\text{FN}}(\tau,t)$ becomes acceptably low. In a symmetric test, by contrast, we treat false positives and false negatives as equally costly and choose the threshold $\tau$ that minimizes the average error. This is equivalent to the Neyman–Pearson criterion with equal prior probabilities and costs for $H_0$ and $H_1$. Effectively, we find $\tau$ that minimizes the mean error probability 
\begin{equation} 
p_{\text{mean}}(t) \;=\; \min_{\tau}\; \frac{p_{\text{FP}}(\tau) + p_{\text{FN}}(\tau,t)}{2}\,. \tag{5}
\end{equation}
Using the measured values of $\eta(t)$ and $\sigma_{\eta(t)}$ for our coherent and squeezed datasets (as obtained from the absorbance analysis and known noise levels), we computed the above error probabilities as a function of time. Figure \ref{fig:eta_paper} provides an intuitive visual comparison of the two probes, while Fig.\,\ref{fig:func_prob_paper} presents the corresponding quantitative results from the hypothesis-testing model. Figure~\ref{fig:eta_paper} plots the transmissivity versus time in the early growth phase, comparing the classical and quantum measurements directly to the blank baseline. 
We fitted the initial portion of each $\eta(t)$ curve with a simple polynomial model $\eta(t) \approx \eta_{bk} - c\,t^2 + d\,t^3$ to guide the eye. The fitted coefficients were nearly identical (differences < 0.5 \%), confirming that both probes measured the same underlying growth kinetics. The earliest detection time can be estimated from Fig.~\ref{fig:eta_paper} by the point at which the growth curve falls a certain amount below the blank. For instance, using a 1$\sigma$ criterion (when the mean $\eta(t)$ is one standard deviation below $\eta_{bk}$), the squeezed-light measurement (red) reaches this threshold roughly 30 minutes earlier than the coherent-light measurement (blue). This preliminary estimate is corroborated and refined by the formal error analysis in Fig.~\ref{fig:func_prob_paper}. 

Figure~\ref{fig:func_prob_paper} presents the error probabilities for distinguishing growth vs.\ no-growth as a function of time, for the classical (blue curves) and quantum (red curves) probes. In Fig.~\ref{fig:func_prob_paper}(a), we show the minimum mean error probability $p_{\text{mean}}(t)$ (symmetric test) at each time. The curves start at 50\% error at $t=0$ (no information), and decrease as growth signals emerge. The red curve drops faster than the blue, crossing below, for example, 5\% mean error at an earlier time. In Fig.~\ref{fig:func_prob_paper}(b), we plot the false-negative probability $p_{\text{FN}}(t)$ under an asymmetric test where the decision threshold $\tau$ is chosen to fix a low false-positive rate (here we chose $p_{\text{FP}} \approx 1\%$ for illustration). Again, the quantum-enhanced measurement outperforms the conventional scheme: at a given time $t$, $p_{\text{FN}}$ is significantly lower for the squeezed probe than for the coherent probe. Equivalently, to reach the same low missed-detection probability, the squeezed-light setup requires less time. For example, in our data $p_{\text{FN}}$ drops below 10\% about one hour earlier with squeezed light than with coherent light (for the chosen false-alarm level). Across both the symmetric and asymmetric testing scenarios, squeezed light enabled detection of bacterial growth roughly 30 minutes sooner than the traditional shot-noise-limited approach, while maintaining comparable confidence against false alarms. These results quantitatively confirm the advantage of quantum-enhanced measurements for early biosensing.

\section{Conclusions}

We have demonstrated that quantum-enhanced optical measurements can accelerate the detection of bacterial growth, a result that might have implications for biosensing and diagnostics. Biological sensing benefits from quantum techniques because the low-light operation of quantum measurements inherently aligns with the need for non-invasiveness in probing living systems. In our experiments, a squeezed-light photometer detected the onset of Escherichia coli growth up to 30 minutes earlier than an equivalent classical photometer at the same power, highlighting how a quantum resource translates into a practical sensing advantage. This study provides the first experimental proof-of-concept that quantum noise reduction can expedite the detection of bacterial growth in real time. 

Our findings open the door to a new class of quantum biosensors that outperform classical limits without increasing illumination intensity. The approach can be generalized to other scenarios where early detection of weak optical signals is vital – for example, detecting low concentrations of pathogens or rare cells, monitoring small changes in blood or tissue samples, or real-time diagnostics in microfluidic devices. Future developments could explore quantum-enhanced discrimination between different bacterial species or strains (by extending the technique to measure spectral signatures or multiple parameters), as well as integration of squeezed-light sources into compact, user-friendly devices for field applications. We envision that quantum-enhanced measurement techniques, alongside advancements in quantum light sources and detectors, could become powerful tools in biological and medical diagnostics, enabling faster and more sensitive detection.
\section*{Funding} The authors acknowledge support from the Novo Nordisk Foundation (CBQS, NNF24SA0088433), the Danish National Research Foundation (bigQ, DNRF142), the NNF NERD grant (NNF20OC0061673) and the European Union's Horizon 2020 research and innovation program under grant agreement No 862644 (QUARTET).


\appendix
\section*{Supplementary Material}
\section{\label{sec:exp-setup} Experimental setup}
\begin{figure*}[htpb]
\centering
\includegraphics[width =\textwidth]{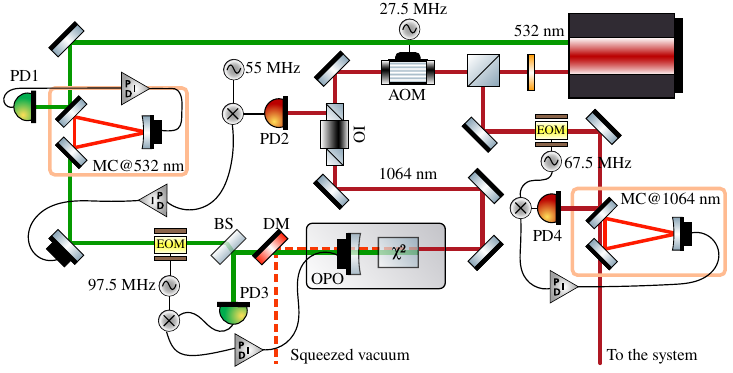}
\caption{Experimental setup. From right to left. Main laser source outputting \SI{1064}{\nano\meter} and \SI{1064}{\nano\meter} optical beams. MC@532 to clean the mode-shape of the \SI{532}{\nano\meter} light generated by the laser system. PD2 plus PID systems are used to phase-locking the phase between pump and seed beams. PD3 plus PID systems are used to stabilize the cavity length of the OPO. OPO is the source that generates vacuum-squeezed states of light. MC@1064 to filter and mode-shape the optical at 1064 nm that will be used as local oscillator in the system. AOM: Acoustic Optic Modulator. OI: Optical Isolator. MC: Mode cleaner cavity. PDs: resonant photo-detectors. EOM: electro-optical modulator. OPO: Optical Parametric Oscillator. BS: beam splitter, DM: dichroic mirror, PID: control system, proportional, integral, and derivative.}
\label{fig:OPO_system}
\end{figure*}

A schematic representation of the experimental setup for the generation of squeezed light is shown in Figure~\ref{fig:OPO_system}. The source of squeezed states of light is a double resonant optical parametric oscillator (OPO) consisting of a periodically poled potassium titanyl phosphate (PPKTP) crystal and a hemispheric coupling mirror. The crystal has \SI{10}{\milli\meter} x \SI{2}{\milli\meter} x \SI{1}{\milli\meter} size, it has two planar facets. The facet used as the end-mirror of the cavity has a high reflectivity coating  ($R>99.9 \%$ at \SI{532}{\nano\meter} and $R>99.95 \%$ at \SI{1064}{\nano\meter}), and the other end-facet has an anti-reflection (AR) coating for both wavelengths. The coupling mirror is attached to a piezo transducer and has a reflectivity of $97.5\%$ at \SI{532}{\nano\meter} and R = $89 \%$ at \SI{1064}{\nano\meter}. The concave mirror has a radius of curvate (ROC) of \SI{-20}{\milli\meter} and is placed around \SI{13}{\milli\meter} from the crystal. The cavity has a full-width-half-maximum bandwidth of around \SI{66}{\mega\hertz}. For achieving double resonant is necessary to change the temperature of the crystal so a peltier element is attached to the crystal. The cavity is stabilized using a Pound-Drever-Hall (PDH) technique by means of a \SI{97.5}{\mega\hertz} phase modulation in the pump beam path.

The main laser source is a \SI{1064}{\nano\meter} Nd:YAG laser from Coherent with \SI{400}{\milli\watt} output power. The \SI{1064}{\nano\meter} beam is split into two paths: one path is divided for use as the pumping beam path and an optical path for the coherent control beam, called the seed beam. The other path is used as the local oscillator for characterizing the amount of squeezing generated in the system or as the local oscillator of the probe beam that will be used to perform the measurements shown in this paper.

The main laser source has an internal Second Harmonic Generation (SHG) unity, a built-in cavity with a lithium niobate (LiNbO3) crystal is optically pumped at \SI{1064}{\nano\meter} to generate light at \SI{532}{\nano\meter}. The \SI{532}{\nano\meter} is filtered by means of a triangular-shaped traveling-wave mode cleaning cavity (MC532nm) before pumping the OPO cavity. The MC532nm is phase-locked using a Pound-Drever-Hall (PDH) technique using the internal \SI{12}{\mega\hertz} phase modulation of the laser.  

The seed beam is used to implement a coherent-locking technique \cite{vahlbruch_coherent_2006} to stabilize the relative phase between the pump beam and the seed beam. The seed beam is frequency-shifted using an acoustic optical modulator (AOM) at \SI{27.5}{\mega\hertz}, and it enters the OPO cavity through the HR side. The up-shifted seed beam interacts with the pump field through difference-frequency generation. The reflected light is detected at a resonant photodetector at \SI{1064}{\nano\meter} (PD2) and down-mixed at \SI{55}{\mega\hertz} to generate an error signal and stabilize the relative phase between the pump and the seed beams. 

\begin{figure}[htpb]
\centering
\includegraphics[width =1\linewidth]{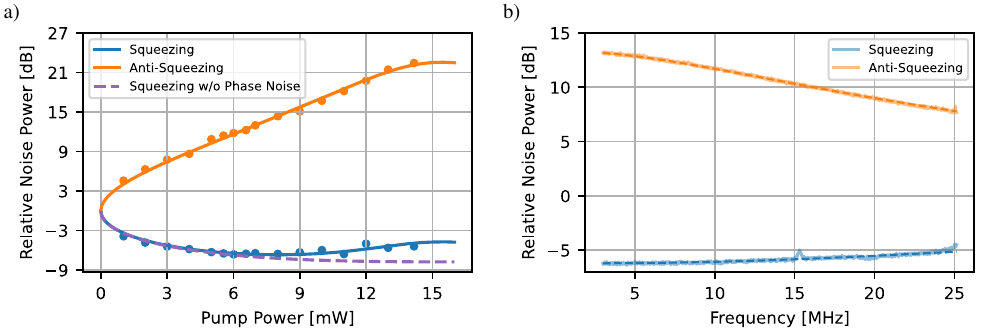}
\caption{Squeezing and anti-squeezing in terms of a) pump power in mW at \SI{5}{\mega\hertz} analysis frequency. b) the analysis frequency in MHz for a \SI{12}{\milli\watt} optical pump power.}
\label{fig:SQZ_paper}
\end{figure}

Before sending the squeezed beam to the actual experiment, we performed a characterization of the degree of squeezing in the system using a homodyne detection scheme. The OPO produced –6.19 dB of vacuum-squeezed light at \SI{1064}{\nano\meter} measured at a sideband frequency of \SI{10}{\mega\hertz}. Figure~\ref{fig:SQZ_paper} shows the amount of squeezing and anti-squeezing produced in the system at Fig.\ref{fig:SQZ_paper}a) \SI{5}{\mega\hertz} in terms of the optical pump power and in Fig.\ref{fig:SQZ_paper}b) in terms of the frequency when pumping with \SI{12}{\milli\watt}.

\section{Growth curve using a commercial spectrophotometer}\label{section_II_commercial_spectrophotometer}

\begin{figure}[htpb]
\centering
\includegraphics[width =1\linewidth]{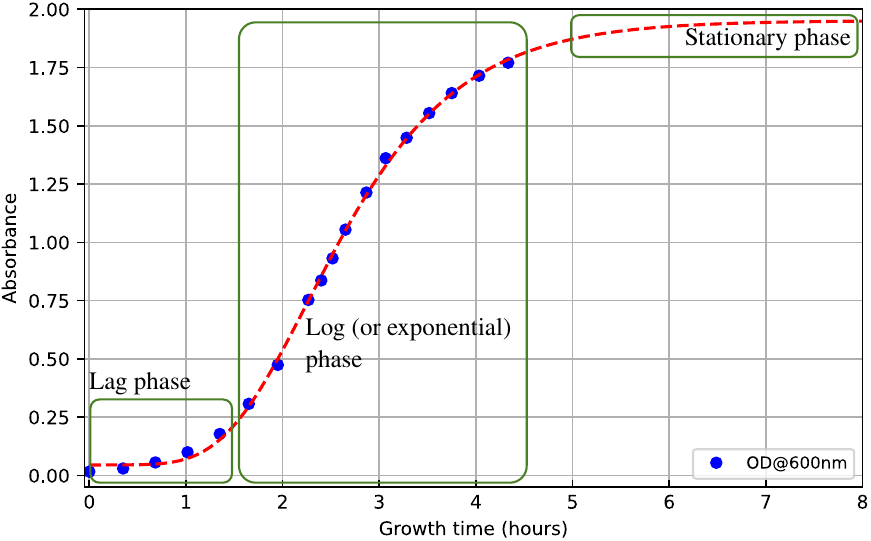}
\caption{Growth curve of {\it E.\ coli}. Absorbance measurements were done at \SI{600}{\nano\meter}. The three main phases are shown: lag phase, log (or exponential) phase, and stationary phase.}
\label{Fig_gc_Ecoli_600nm}
\end{figure}

Bacterial growth is traditionally measured using spectrophotometer systems, which are optical instruments that measure the optical density of a sample. Typically a growth curve has four main phases: lag phase, exponential phase, stationary phase, and death phase. These phases are shown in Figure~\ref{Fig_gc_Ecoli_600nm}. 

\begin{figure}[htpb]
\centering
\includegraphics[width =1\linewidth]{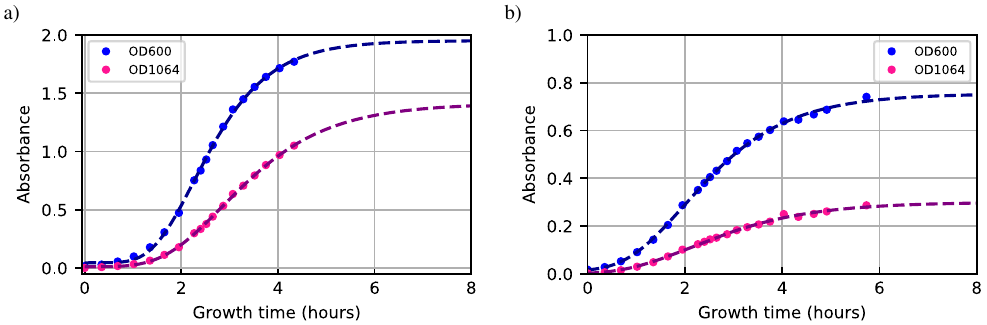}
\caption{{\it E.\ coli} growth curve. Absorption measurements were taken at \SI{600}{\nano\meter} and \SI{1064}{\nano\meter}. Cell culture was growing in a) a \SI{1000}{\milli\liter} Erlenmeyer glass inside the bench-top orbital shake incubator. b) a \SI{3}{\milli\liter} cuvette in the mini-incubator. Both systems were operating at \SI{37}{\celsius} and 250 $rpm$. Absorbance measurements were done using two different wavelengths: in blue -  \SI{600}{\nano\meter} and in pink - \SI{1064}{\nano\meter}.}
\label{Fig_spectrophotometer_Ecoli}
\end{figure}

The protocol for preparing the bacteria culture was the following: from a Luria-Bertani (LB) agar plate, a single colony of bacteria was inoculated into \SI{3}{\milli\liter} pre-warmed LB broth media in a test tube. The test tube was placed in a bench-top orbital shaker-incubator (inc1, compact shaker-incubator ES-20 from Grant-bio) set to \SI{37}{\celsius} and 250 $rpm$ overnight (O/N). A 10 times dilution sample was prepared with the O/N culture, and the optical density (OD) was measured at \SI{600}{\nano\meter} (OD600) using a visible spectrophotometer (VWR PV4). From the OD measurement, we estimated the volume of O/N culture necessary for growing the cell culture in two different sample containers (and incubators): a \SI{1000}{\milli\liter} Erlenmeyer glass and a \SI{3}{\milli\liter} cuvette. The results can be seen in Figure~\ref{Fig_spectrophotometer_Ecoli}. 

In Fig. \ref{Fig_spectrophotometer_Ecoli}a) the cell culture grows in a \SI{1000}{\milli\liter} Erlenmeyer glass inside the bench-top orbital shake incubator kept at \SI{37}{\celsius} and shaking at 250 $rpm$. In Fig. \ref{Fig_spectrophotometer_Ecoli}b) the cell culture grows in a \SI{3}{\milli\liter} cuvette in the mini-incubator (inc2, air-cooled, Peltier-controlled single cuvette holder from Mountain Photonics (QNW qX2)). Both incubators were operating at \SI{37}{\celsius} and 250 $rpm$. Both samples are measured in the visible spectrophotometer every 20 minutes. The spectrophotometer performs measurements in a range of \SIrange{320}{1100}{\nano\meter}, and for this work, we did measurements at \SI{600}{\nano\meter} and \SI{1064}{\nano\meter}. The OD measurement at \SI{600}{\nano\meter} is commonly used to determine the stage of growth of a bacterial culture. The OD measurement at \SI{1064}{\nano\meter} allows us to compare the quantum-enhanced measurement results with measurements done using the commercial system. We notice from Fig. \ref{Fig_spectrophotometer_Ecoli}a) that the absorption values at \SI{600}{\nano\meter} are larger than the measurement done at \SI{1064}{\nano\meter}, which is expected as the absorption profile of {\it E.\ coli} decreases when the wavelength increases \cite{curtis_wr_improving_nodate}. Fig. \ref{Fig_spectrophotometer_Ecoli}b) shows smaller absorbance values for the sample growing in the mini-incubator, possibly due to the effect of the decrease of the oxygen rate exchange when growing {\it E.\ coli} using a mini-incubator (inc2). 

From the measurement data the analytical form of the growth curves is analyzed using the Gompertz function \cite{zwietering_modeling_1990-1} $ A = a \exp \left\{-\exp\left[\frac{\mu e}{a}(\theta - t)+1\right]\right\} + A_{bk}$. Where $a$ is the asymptotic absorbance, $\mu$ is the rate of growth in the linear region, $\theta$ is the lag time, and $A_{bk}$ accounts for nonzero mean absorbance of the blank sample (fresh growth media). The Gompertz parameters are shown in table \ref{tab:gompertz_fig5}.

\subsection{Analysis of the Gompertz function parameters}\label{subsection_gompertz_function}
The Gompertz function in terms of parameters that are microbiologically relevant is given by:

\begin{equation}
    A = a \exp \left\{-\exp\left[\frac{\mu e}{a}(\theta - t)+1\right]\right\} + A_{bk}
    \label{func_gompertz}
\end{equation}
where $a$ is the asymptotic absorbance when $t \rightarrow \infty$, $\mu$ is the rate of growth in the linear region, $\theta$ is the lag time, and $A_{bk}$ accounts for absorbance of the blank sample. The absorbance $A$ is defined as:

\begin{equation}
    A = \log_{10}\left(\frac{I_{blank}}{I_{growth}}\right)
\end{equation}

A growth curve has three main phases: the lag phase, the exponential phase, and the stationary phase. During the exponential phase, we compute the doubling time of the bacterial growth.

During the lag phase, the number of cells remains unchanged, and in this phase, the cells start to adjust to the medium. The length of this phase depends on the number of cells inoculated in the medium, the time to recover from physical damage, and the time required to synthesize enzymes.

In the log or exponential phase, the cells multiply at the fastest rate, and the population doubles at each generation's time.

During the stationary phase, the nutrients in the medium exhaust after they are completely utilized, and the cells stop dividing.

An important biological parameter is the doubling time of the bacterial growth. For estimating the doubling time, first, we compute the maximum growth rate, named $\mu$, given by the slope of the curve when the organisms grow exponentially. By fitting the Gompertz function to the data set, we can determine $\mu$, $\theta$, $A_{bk}$, and $a$.

Considering that we know the maximum growth rate $\mu$, we compute the doubling time in the following way: In the exponential region, we can model the area as $f(t) = \alpha e^{\beta t}$. When the cells grow exponentially, the region can be described by a linear growth phase. 

\begin{equation}
    \log_{10}(f(t)) = \log_{10}(\alpha) + \log_{10}(e)\beta t
\end{equation}
This phase, in terms of the Gompertz function parameters, is a tangent line through the inflection point:

\begin{eqnarray}
y(t)& =& \mu (t-t_{i}) + \frac{a}{e} , t_{i} = (\theta + \frac{a}{\mu e}) \ \ \ \rightarrow\ \ \  \nonumber
\\
y(t) &=& \mu(t-\theta).
\end{eqnarray}

This implies that $y(t) = \log_{10}(f(t))$, $-\mu \theta = \log_{10}(\alpha) $ and $\mu = \beta \log_{10}(e)$.

We want to know when the OD doubles in the exponential phase. We will name the doubling time as $\Delta$:

\begin{eqnarray}
  f(t+\Delta) &=& 2f(t) \ \ \ \rightarrow\ \ \  \alpha e^{\beta.(t+ \Delta)} = 2\alpha e^{\beta.t} \ \ \ \rightarrow\ \ \  \nonumber 
 \\
  e^{\beta.\Delta} &=& 2 \ \ \ \rightarrow\ \ \    \beta.\Delta = \ln(2) \ \ \ \rightarrow\ \ \  \nonumber 
\\
\Delta &=& \frac{\ln(2)\log_{10}(e)}{\mu}       
\end{eqnarray}

So, we can plot a linear function in the exponential growth phase with parameters $\mu$ and $\theta$ determined in the Gompertz function. Given the growth rate $\mu$, we can estimate the doubling time of this bacterial population. 

\subsection{Table with parameters from the Gompertz curves above}
\begin{figure}[htpb]
\centering
\includegraphics[width =1\linewidth]{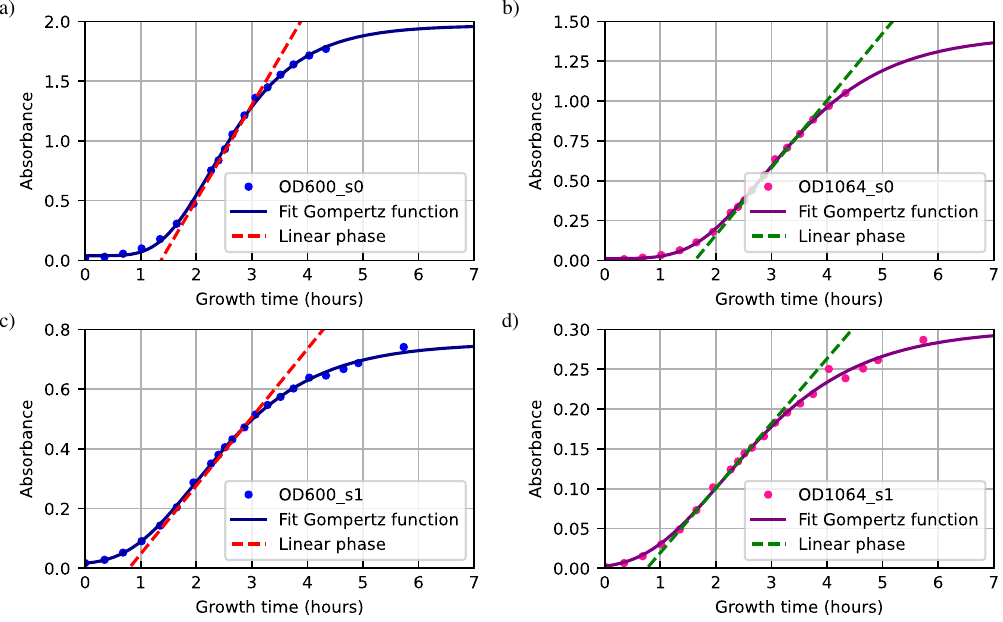}
\caption{{\it E.\ coli} growth curve. Cell culture was prepared in a \SI{1000}{\milli\liter} Erlenmeyer glass kept inside the bench-top orbital shake incubator. Absorption measurements were taken at a) \SI{600}{\nano\meter}.  b) \SI{1064}{\nano\meter}. Cell culture was prepared in a \SI{3}{\milli\liter} cuvette kept inside the mini-incubator. Absorption measurements were taken at c) \SI{600}{\nano\meter}. d) \SI{1064}{\nano\meter}. The cells were growing under \SI{37}{\celsius} and shaking at 250 $rpm$ in both containers.}
\label{Fig_Growth_Curve_OD600_OD1064_samples}
\end{figure}

For reference curves of the growth of {\it E.\ coli} in the quantum-enhanced setup, we made a set of characterizations using a commercial spectrophotometer as described at the beginning of section \ref{section_II_commercial_spectrophotometer}. We characterized the bacterial growth using two different sample containers, a \SI{1000}{\milli\liter} Erlenmeyer glass in a bench-top shake incubator and a \SI{3}{\milli\liter} cuvette placed inside a mini-incubator. We measured the OD of the sample growing in each container using \SI{1064}{\nano\meter} and \SI{600}{\nano\meter} wavelengths. This configuration choice was made so we could first make sure that the sample we inoculated in the growth medium was growing in accordance with the literature when using a bench-top incubator, a \SI{1000}{\milli\liter} Erlenmeyer glass, and \SI{600}{\nano\meter} wavelength for characterizing the OD. The measurements performed in the cell culture growing in the mini-incubator using \SI{1064}{\nano\meter} light was used as a classical reference for the quantum-enhanced spectrophotometer, in which the mini-incubator was used during the growth characterization. The classical results are shown in Fig. \ref{Fig_Growth_Curve_OD600_OD1064_samples}.

In each curve, we fit the function \ref{func_gompertz} and obtained the parameters  $\mu$, $\theta$, $A_{bk}$, and $a$. These parameters are shown in table \ref{tab:gompertz_fig5}. Given we have parameters $\mu$ and $\theta$, we fit a linear curve in the exponential region to show the linear behavior of the system, as described in section \ref{subsection_gompertz_function}. 

\begin{table}[htbp]
\centering
\caption{\bf Parameters of the Gompertz function}
\begin{tabular}{|c|c|c|c|c|}
\hline
Fig. \ref{Fig_Growth_Curve_OD600_OD1064_samples} & $a$ & $\mu [1/h]$ & $\theta [h]$ & $A_{bk}$ \\
\hline
 \multicolumn{5}{|c|}{OD1064} \\
 \hline
a)& $1.40 \pm 0.03$ & $0.422 \pm 0.005$ & $1.62 \pm 0.02$ & $0.010 \pm 0.004$ \\ 
b) & $0.30 \pm 0.01$ & $0.081 \pm 0.003$ & $0.76 \pm 0.10$ & $1.85 e^{-10} \pm 0.0002$ \\
\hline
 \multicolumn{5}{|c|}{OD600} \\
 \hline
c) & $1.93  \pm 0.03 $ & $0.80 \pm 0.01$ & $ 1.37 \pm 0.03$ & $0.04 \pm 0.01$ \\
d) &  $0.74 \pm 0.01$ & $0.229 \pm 0.004$ & $ 0.79 \pm 0.06$ & $ 0.013 \pm 0.007$  \\
\hline
\end{tabular}
  \label{tab:gompertz_fig5}
\end{table}

We need each data set's $\mu$ parameter to compute the doubling time. Table \ref{tab:doubling_time} shows the doubling time for each configuration in \ref{Fig_Growth_Curve_OD600_OD1064_samples}.

\begin{table}[htbp]
\centering
\caption{\bf Parameters of the Gompertz function}
\begin{tabular}{|c|c|c|}
\hline
Fig. \ref{Fig_Growth_Curve_OD600_OD1064_samples} & $\mu [1/h]$ & $\Delta [min]$  \\
\hline
\multicolumn{3}{|c|}{OD600} \\
\hline
a) & $0.80 \pm 0.01$ & $22.70$ \\
c) & $0.229 \pm 0.004$ & $78.94$ \\
\hline
\multicolumn{3}{|c|}{OD1064} \\
 \hline
b) & $0.422 \pm 0.005 $ & $42.78$ \\
d) & $0.081 \pm 0.003$ & $222.30$ \\
\hline
\end{tabular}
 \label{tab:doubling_time}
\end{table}

The doubling time measured at OD600 for the sample growing in the Erlenmeyer glass in the bench-top incubator follows the usual behavior in the literature, a doubling time of around 20 min for {\it E.\ coli} under optimal conditions\cite{tuttle_growth_2021, raynaud_molecular_2011}. When growing {\it E.\ coli} in the mini-incubator, longer doubling times are expected, possibly due to the effect of the decrease of the oxygen rate exchange. The absorption profile of {\it E.\ coli} changes with the wavelength chosen to measure the OD. Absorption values measured at \SI{600}{\nano\meter} are larger than at \SI{1064}{\nano\meter}. On the one hand, the responsivity of the photodetector present in the spectrophotometer decreases with the wavelength. On the other hand, the light's intensity decreases with the wavelength's fourth power due to Rayleigh scattering \cite{curtis_wr_improving_nodate}.

\section{Complementary experimental data - quantum-enhanced setup}

\subsection{Raw data set}

\begin{figure}[htpb]
\centering
\includegraphics[width =1\linewidth]{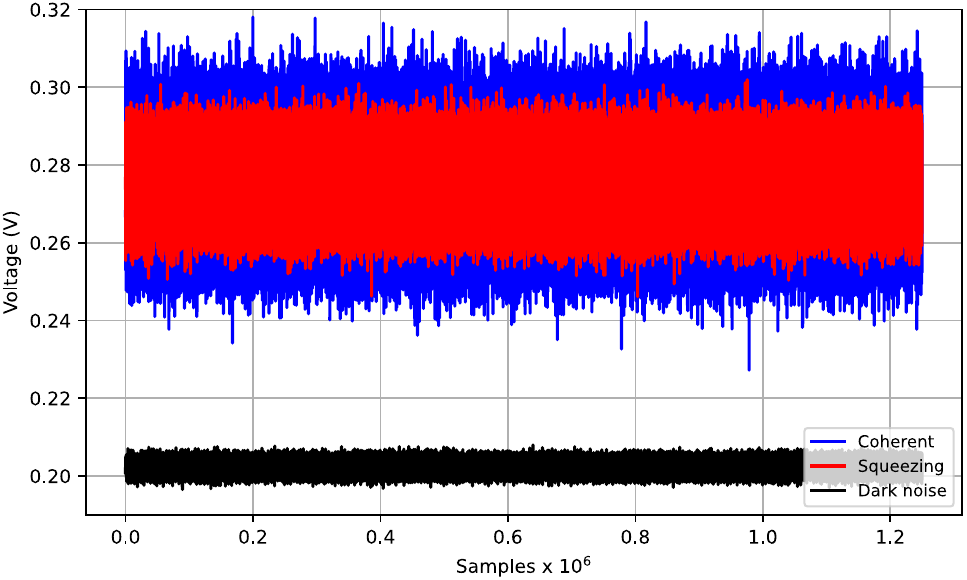}
\caption{Example of a raw data trace. Measurement in - red squeezed state and - blue coherent state of light.}
\label{fig_raw_data}
\end{figure}

A typical 5-second long-time trace representing one measurement data set is shown in Fig.~\ref{fig_raw_data}. A clear noise reduction is visible when comparing the squeezed probe with the coherent one. The dark noise - measurement without light - is also shown in the figure. For each data point, for example, in ref{}, we averaged the 5-second-long time trace to obtain the mean value. 

\subsection{Probe optical power fluctuations - monitoring}
To account for optical power fluctuations from the \SI{1064}{\nano\meter} laser system, the optical power of the carrier of the coherent displacement beam was acquired simultaneously with the time trace for the absorption measurement. The power of the coherent beam was controlled by a half-waveplate and a polarizing beam splitter and monitored with a power meter (see Fig ~\ref{fig:Experimental_setup} in the main paper).

\begin{figure}[htpb]
\centering
\includegraphics[width =1\linewidth]{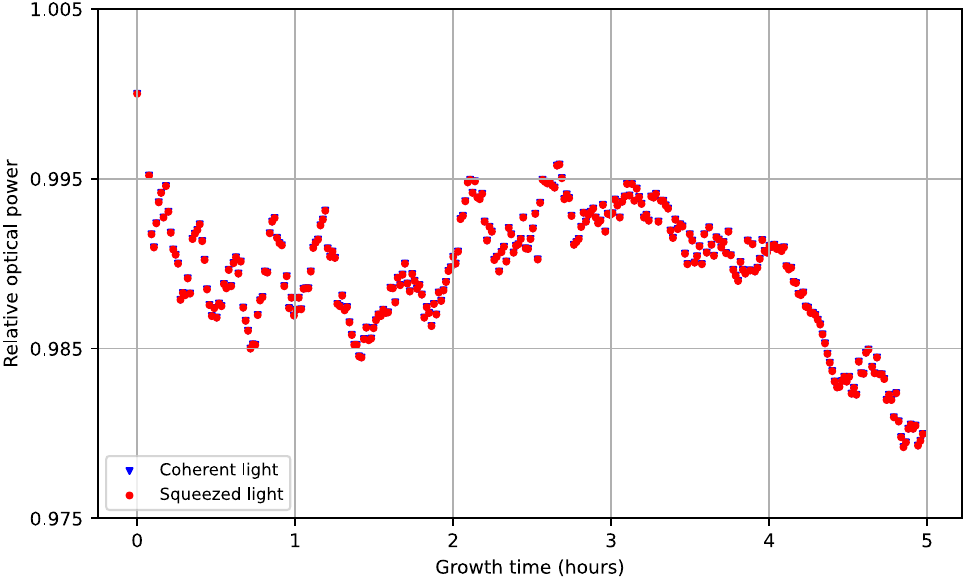}
\caption{\SI{1064}{\nano\meter} laser optical power fluctuations. The values were normalized by the first optical power measurement.}
\label{fig_power_fluctuation}
\end{figure}

Figure~\ref{fig_power_fluctuation} shows the fluctuations in the optical power of the \SI{1064}{\nano\meter} laser over the 5-hour measurement duration. We normalized all the measurements in the main paper to compensate for these optical power fluctuations.
\section{Truncated Gaussian Function}

The truncated function has the following PDF (probability density function):

\begin{equation}
    \Psi(\overline{\mu}, \overline{\sigma}, a, b; x)= \frac{\phi(\overline{\mu}, \overline{\sigma}^{2}; x)}{[\Phi(\overline{\mu}, \overline{\sigma}^{2}; b)-\Phi(\overline{\mu}, \overline{\sigma}^{2}; a)]}
\end{equation}

Where $\overline{\mu}$ and $\overline{\sigma}$ are the mean and variance of the general normal distribution, $a$ and $b$ are the parameters that define the truncation interval, and $\phi(\overline{\mu}, \overline{\sigma}^{2}; x)$ is the PDF of the general normal distribution:

\begin{equation}
    \phi(\overline{\mu}, \overline{\sigma}^{2}; x) = \frac{1}{\overline{\sigma}\sqrt{2\pi}}\exp\left[\frac{-(x-\overline{\mu})^{2}}{2\overline{\sigma}^{2}}\right]
\end{equation}

and $\Phi(\overline{\mu}, \overline{\sigma}^{2}; x)$ is the CDF of the general normal distribution:

\begin{equation}
    \Phi(\overline{\mu}, \overline{\sigma}^{2}; x) = \frac{1}{\overline{\sigma}\sqrt{2\pi}} \int_{-\infty}^{\infty} \exp\left[\frac{-(t-\overline{\mu})^{2}}{2\overline{\sigma}^{2}}\right]dt
\end{equation}

For the problem we are working on in this paper, the truncation parameters a and b are $[0, \infty)$.  So we are looking for a PDF as:

\begin{eqnarray}
    \Phi(\overline{\mu}, \overline{\sigma}^{2}; b) &=& \frac{1}{2}\left(1+\operatorname{erf}\left(\frac{b-\overline{\mu}}{\sqrt{2}\overline{\sigma}}\right)\right) \nonumber \\ 
    if \ \ b  &\rightarrow& \infty \nonumber \\
    \Phi(\overline{\mu}, \overline{\sigma}^{2}; b) &\rightarrow& 1 \\
    \Phi(\overline{\mu}, \overline{\sigma}^{2}; a) &=& \frac{1}{2}\left(1+\operatorname{erf}\left(\frac{a-\overline{\mu}}{\sqrt{2}\overline{\sigma}}\right)\right) \nonumber \\ 
    if \ \ a  &\rightarrow& 0 \nonumber \\
    \Phi(\overline{\mu}, \overline{\sigma}^{2}; b) &\rightarrow& \frac{1}{2}\left(1+\operatorname{erf}\left(\frac{-\overline{\mu}}{\sqrt{2}\overline{\sigma}}\right)\right)
\end{eqnarray}

Resulting in:

\begin{equation}
    \Psi(\overline{\mu}, \overline{\sigma}, 0, b; x)= 2\frac{\phi(\overline{\mu}, \overline{\sigma}^{2}; x)}{1-\operatorname{erf}\left(\frac{-\overline{\mu}}{\sqrt{2}\overline{\sigma}}\right)}
\end{equation}

Now let's compute the mean and the variance of the truncated normal distribution. Defining $\alpha = \frac{a-\overline{\mu}}{\overline{\sigma}}$ and $\beta = \frac{b-\overline{\mu}}{\overline{\sigma}}$, the mean value can be computed as (cite Johson): 

\begin{equation}
    \mu = \overline{\mu} - \overline{\sigma}\frac{\phi(0, 1; \beta)-\phi(0, 1; \alpha)}{\Phi(0, 1; \beta)-\Phi(0, 1; \alpha)}
\end{equation}

Evaluating the individual terms of the equation above, we have:

\begin{eqnarray}
    \phi(0, 1; \beta) = \frac{1}{\sqrt{2\pi}}\exp\left(-\frac{\beta^{2}}{2\sigma^{2}}\right) \nonumber \\
    \beta \rightarrow \infty \Rightarrow \phi(0, 1; \beta) \rightarrow 0 \nonumber \\
    \phi(0, 1; \alpha) = \frac{1}{\sqrt{2\pi}}\exp\left[-\left(\frac{-\mu}{\sqrt{2}\sigma}\right)^{2}\right] \nonumber \\
\end{eqnarray}
Resulting in the mean value of the truncated Gaussian distribution:

\begin{equation}
    \mu = \overline{\mu} - 2\frac{\phi(0, 1; -\frac{\overline{\mu}}{\overline{\sigma}})}{1-\operatorname{erf}\left(-\frac{\overline{\mu}}{\overline{\sigma}\sqrt{2}}\right)}\overline{\sigma}
\end{equation}

We will rewrite the above equation as:

\begin{eqnarray}
    \mu = \overline{\mu} + g(\omega) \overline{\sigma}
\end{eqnarray}

where $\omega = - \frac{\mu}{\sigma}$ and $g(\omega) = 2\frac{\mathcal{N(\omega)}}{1-\operatorname{erf}(\omega/\sqrt{2})}$.

For the variance, we will follow the same approach. Thus we get:

\begin{eqnarray}
 \sigma^{2} &=& \overline{\sigma}^{2}\left[1 - \left(\frac{\beta \phi(0,1;\beta) - \alpha \phi(0,1;\alpha)}{\Phi(0,1;\beta)-\Phi(0,1;\alpha)}\right)  \right. \nonumber
\\
&-& \left. \left(\frac{\phi(0,1;\beta)-\phi(0,1;\alpha)}{\Phi(0,1;\beta)-\Phi(0,1;\alpha)}\right)^{2}\right]
\end{eqnarray}

Rewriting in terms of $g(\omega)$:

\begin{equation}
    \sigma^{2} = \overline{\sigma}^{2}(1 + \omega g(\omega)-g(\omega)^{2}).
\end{equation}

The mean and variance of the truncated normal distribution can be seen as a perturbation in the mean and variance values of the general normal distribution. Figure (figure to be generated) shows the PDF and the CDF of a truncated normal distribution using the following parameters, similar to the ones used in the main paper.

\bibliographystyle{unsrt}
\bibliography{reference_QEA_paper_DTU}

\begin{thebibliography}{10}

\bibitem{zhao2024}
Xinyi Zhao, Abhijnan Bhat, Christine O’Connor, James Curtin, Baljit Singh,
  and Furong Tian.
\newblock Review of detection limits for various techniques for bacterial
  detection in food samples.
\newblock {\em Nanomaterials}, 14(10):855, 2024.

\bibitem{marro2022}
Florian~C Marro, Fr{\'e}d{\'e}ric Laurent, J{\'e}r{\^o}me Josse, and Ariel~J
  Blocker.
\newblock Methods to monitor bacterial growth and replicative rates at the
  single-cell level.
\newblock {\em FEMS Microbiology Reviews}, 46(6):fuac030, 2022.

\bibitem{Mcgoverin2021}
C.~Mcgoverin, C.~Steed, A.~Esan, J.~Robertson, S.~Swift, and F.~Vanholsbeeck.
\newblock Optical methods for bacterial detection and characterization.
\newblock {\em APL Photonics}, 6, 2021.

\bibitem{Hu2017}
J.~Hu and P.~W. Bohn.
\newblock Optical biosensing of bacteria and bacterial communities.
\newblock {\em Journal of Analysis and Testing}, 1, 2017.

\bibitem{Locke2020}
A.~Locke, S.M. Fitzgerald, and A.~Mahadevan-Jansen.
\newblock Advances in optical detection of human-associated pathogenic
  bacteria.
\newblock {\em Molecules}, 25, 2020.

\bibitem{Dubourg2018}
G.~Dubourg, B.~Lamy, and R.~Ruimy.
\newblock Rapid phenotypic methods to improve the diagnosis of bacterial
  bloodstream infections: meeting the challenge to reduce the time to result.
\newblock {\em Clinical Microbiology and Infection}, 24:935--943, 2018.

\bibitem{fu_characterization_2006}
Yan Fu, Haifeng Wang, Riyi Shi, and Ji-Xin Cheng.
\newblock Characterization of photodamage in coherent anti-stokes raman
  scattering microscopy.
\newblock {\em Optics Express}, 14(9):3942, 2006.

\bibitem{galli_intrinsic_2014}
Roberta Galli, Ortrud Uckermann, Elisabeth~F. Andresen, Kathrin~D. Geiger,
  Edmund Koch, Gabriele Schackert, Gerald Steiner, and Matthias Kirsch.
\newblock Intrinsic indicator of photodamage during label-free multiphoton
  microscopy of cells and tissues.
\newblock {\em {PLoS} {ONE}}, 9(10):e110295, 2014.

\bibitem{talone_phototoxicity_2021}
B.~Talone, M.~Bazzarelli, A.~Schirato, F.~Dello~Vicario, D.~Viola,
  E.~Jacchetti, M.~Bregonzio, M.~T. Raimondi, G.~Cerullo, and D.~Polli.
\newblock Phototoxicity induced in living {HeLa} cells by focused femtosecond
  laser pulses: a data-driven approach.
\newblock {\em Biomedical Optics Express}, 12(12):7886, 2021.

\bibitem{Belkum2020}
A.~v. Belkum, C.~D. Burnham, J.~W.~A. Rossen, F.~Mallard, O.~Rochas, and W.~M.
  Dunne.
\newblock Innovative and rapid antimicrobial susceptibility testing systems.
\newblock {\em Nature Reviews Microbiology}, 18:299--311, 2020.

\bibitem{brida2010}
G.~Brida, M.~Genovese, and Ruo Berchera.
\newblock Experimental realization of sub-shot-noise quantum imaging.
\newblock {\em Nature Photon}, 4:8227, 2010.

\bibitem{andrade_quantum-enhanced_2020}
Rayssa B.~de Andrade, Hugo Kerdoncuff, Kirstine Berg-Sørensen, Tobias Gehring,
  Mikael Lassen, and Ulrik~L. Andersen.
\newblock Quantum-enhanced continuous-wave stimulated raman scattering
  spectroscopy.
\newblock {\em Optica}, 7(5):470--475, 2020.

\bibitem{xu_stimulated_2022}
Zicong Xu, Kenichi Oguchi, Yoshitaka Taguchi, Yuki Sano, Yu~Miyawaki, Donguk
  Cheon, Kazuhiro Katoh, and Yasuyuki Ozeki.
\newblock Stimulated raman scattering spectroscopy with quantum-enhanced
  balanced detection.
\newblock {\em Optics Express}, 30(11):18589, 2022.

\bibitem{taylor_biological_2013}
Michael~A. Taylor, Jiri Janousek, Vincent Daria, Joachim Knittel, Boris Hage,
  Hans-A. Bachor, and Warwick~P. Bowen.
\newblock Biological measurement beyond the quantum limit.
\newblock {\em Nature Photonics}, 7(3):229--233, 2013.

\bibitem{casacio_quantum-enhanced_2021}
Catxere~A. Casacio, Lars~S. Madsen, Alex Terrasson, Muhammad Waleed, Kai
  Barnscheidt, Boris Hage, Michael~A. Taylor, and Warwick~P. Bowen.
\newblock Quantum-enhanced nonlinear microscopy.
\newblock {\em Nature}, 594(7862):201--206, 2021.

\bibitem{spedalieri_detecting_2020}
Gaetana Spedalieri, Lolita Piersimoni, Omar Laurino, Samuel~L. Braunstein, and
  Stefano Pirandola.
\newblock Detecting and tracking bacteria with quantum light.
\newblock {\em Physical Review Research}, 2(4):043260, 2020.

\bibitem{tjorve_use_2017}
Kathleen M.~C. Tjørve and Even Tjørve.
\newblock The use of gompertz models in growth analyses, and new gompertz-model
  approach: An addition to the unified-richards family.
\newblock {\em {PLOS} {ONE}}, 12(6):1--17, 2017.

\bibitem{mira_estimating_2022}
Portia Mira, Pamela Yeh, and Barry~G. Hall.
\newblock Estimating microbial population data from optical density.
\newblock {\em {PLOS} {ONE}}, 17(10):e0276040, 2022.

\bibitem{krishnamurthi_new_2021}
Venkata~Rao Krishnamurthi, Isabelle~I. Niyonshuti, Jingyi Chen, and Yong Wang.
\newblock A new analysis method for evaluating bacterial growth with microplate
  readers.
\newblock {\em {PLOS} {ONE}}, 16(1):e0245205, 2021.

\bibitem{spedalieri_optimal_2021}
Gaetana Spedalieri and Stefano Pirandola.
\newblock Optimal squeezing for quantum target detection.
\newblock {\em Physical Review Research}, 3(4):L042039, 2021.

\bibitem{vahlbruch_coherent_2006}
Henning Vahlbruch, Simon Chelkowski, Boris Hage, Alexander Franzen, Karsten
  Danzmann, and Roman Schnabel.
\newblock Coherent control of vacuum squeezing in the gravitational-wave
  detection band.
\newblock {\em Physical Review Letters}, 97(1):011101, 2006.

\bibitem{curtis_wr_improving_nodate}
W.R Curtis, B.S Curtis, and J.A Myers.
\newblock Improving accuracy of cell and chromophore concentration measurements
  using optical density.
\newblock {\em BMC Biophys}, 6:4, 2013.

\bibitem{zwietering_modeling_1990-1}
M.~H. Zwietering, I.~Jongenburger, F.~M. Rombouts, and K.~van~'t Riet.
\newblock Modeling of the {Bacterial} {Growth} {Curve}.
\newblock {\em Applied and Environmental Microbiology}, 56(6):1875--1881, 1990.

\bibitem{tuttle_growth_2021}
Amie~R. Tuttle, Nicholas~D. Trahan, and Mike~S. Son.
\newblock Growth and maintenance of \textit{{Escherichia} coli} laboratory
  strains.
\newblock {\em Current Protocols}, 1(1):e20, 2021.

\bibitem{raynaud_molecular_2011}
Céline Raynaud, Jieun Lee, Patricia Sarçabal, Christian Croux, Isabelle
  Meynial-Salles, and Philippe Soucaille.
\newblock Molecular {Characterization} of the {Glycerol}-{Oxidative} {Pathway}
  of {Clostridium} butyricum {VPI} 1718.
\newblock {\em Journal of Bacteriology}, 193(12):3127--3134, 2011.

\end{thebibliography}

\end{document}